# Size Dependence and Ballistic Limits of Thermal Transport in Anisotropic Layered Two-Dimensional Materials

Zuanyi Li,[1,2] Yizhou Liu,[3,4] Lucas Lindsay,[5,†] Yong Xu,[3] Wenhui Duan,[3,4] and Eric Pop[1,*]

[1]*Electrical Engineering, Stanford University, Stanford, California 94305, USA*
[2]*Department of Physics, University of Illinois, Urbana-Champaign, Illinois 61801, USA*
[3]*Dept. of Physics and State Key Laboratory of Low-Dimensional Quantum Physics, Tsinghua University, Beijing 100084, P.R. China*
[4]*Collaborative Innovation Center of Quantum Matter, Beijing 100084, P.R. China*
[5]*Oak Ridge National Laboratory, Oak Ridge, Tennessee 37831, USA*

Layered materials have uncommonly anisotropic thermal properties due to their strong in-plane covalent bonds and weak out-of-plane van der Waals interactions. Here we examine heat flow in graphene (graphite), $h$-BN, MoS$_2$, and WS$_2$ monolayers and bulk films, from diffusive to ballistic limits. We determine the ballistic thermal conductance limit ($G_{ball}$) both in-plane and out-of-plane, based on full phonon dispersions from first-principles calculations. An overall phonon mean free path ($\lambda$) is expressed in terms of $G_{ball}$ and the diffusive thermal conductivity, consistent with kinetic theory if proper averaging of phonon group velocity is used. We obtain a size-dependent thermal conductivity $k(L)$ in agreement with available experiments, and find that $k(L)$ only converges to >90% of the diffusive thermal conductivity for sample sizes $L \geq 16\lambda$, which ranges from ~140 nm for MoS$_2$ cross-plane to ~10 μm for suspended graphene in-plane. These results provide a deeper understanding of microscopic thermal transport, revealing that device scales below which thermal size effects should be taken into account are generally larger than previously thought.

[*]Contact: epop@stanford.edu

[†]This manuscript has been co-authored (L.L.) by UT-Battelle, LLC under Contract No. DE-AC05-00OR22725 with the U.S. Department of Energy. The United States Government retains and the publisher, by accepting the article for publication, acknowledges that the United States Government retains a non-exclusive, paid-up, irrevocable, world-wide license to publish or reproduce the published form of this manuscript, or allow others to do so, for United States Government purposes. The Department of Energy will provide public access to these results of federally sponsored research in accordance with the DOE Public Access Plan (http://energy.gov/downloads/doe-public-access-plan).

Stimulated by extensive studies of graphene [1-3], much interest now exists in the properties of other two-dimensional (2D) layered materials [4-7], such as hexagonal boron nitride (*h*-BN) and transition metal dichalcogenides (TMDs, *e.g.*, $MoS_2$, and $WS_2$). Understanding their thermal properties is important for improving thermally-limited performance in integrated electronics and energy conversion devices [8-11]. The thermal properties of 2D layered materials are unusual and highly anisotropic, due to strong in-plane chemical bonding and weak interlayer van der Waals interactions [1,12]. For instance, bulk graphite has over 300-fold higher in-plane than out-of-plane thermal conductivity at room temperature; in *h*-BN this anisotropy is 200-fold. Like in other materials, when sample dimensions become comparable to the average phonon scattering length or mean free path (MFP) $\lambda$, thermal transport can no longer be described with classical parameters (such as the bulk thermal conductivity) and so-called size effects become important [8].

Simply stated, if the sample size (*i.e.* "length") in the direction of heat flow $L \ll \lambda$, heat conduction enters a ballistic regime (with no scattering) and the thermal conductance approaches a ballistic upper limit, $G_{ball}$, which is only a function of temperature. Interestingly, if the classical relationship between conductance and conductivity is applied ($k = GL/A$) the thermal conductivity becomes size-dependent and approaches zero in the limit of the shortest ballistic samples. The ballistic thermal conductance of carbon nanotubes [13], silicon [14] and graphene in-plane [15] have been recently examined, but these limits are unknown in other 2D (layered) materials, and in their out-of-plane direction. More generally speaking, much less is known about the ballistic limits of thermal transport compared to those of electrical transport [16-19].

If the sample size $L \gg \lambda$, heat conduction is said to be in the classical (Fourier) regime, with the thermal conductivity being independent of sample size. However, the transition from ballistic to diffusive is not well understood, nor is the apparent evolution of thermal conductivity with size, $k(L)$, in this regime. Some theoretical models have predicted that in low-dimensional momentum-conserving systems $k$ will diverge with $L$, scaling as $\sim L^\alpha$ with $\alpha = 0.22$ to $0.5$ for one-dimension (1D) [20-22] and as $\sim \log(L)$ for 2D [22-24]. Although it is natural to assume that the transition from ballistic to diffusive should occur over sample sizes of the order of the phonon MFP, this value is often not estimated consistently in the literature. For example, the Si phonon MFP is ~40 nm at room temperature [25] based on a Debye heat capacity model with an average sound velocity, but ~300 nm when the phonon dispersion is taken into account [26]. Analysis of cumulative thermal conductivity as a function of MFP has suggested that phonons with MFP > 1 μm contribute ~40% to $k$ in Si [27-31], so a "median" MFP of 0.5-1 μm has also been suggested [25,31]. Thus, an improved understanding of size effects and phonon MFP are crucial both in bulk and in atomically thin materials. While a *range* of MFPs provides insight into the spectral distribution of phonon contributions [27-31], providing a *single* MFP gives information in a concise way, being useful to evaluate size effects and transport in practical devices, where the device dimensions are known.

In this work, we study the ballistic-diffusive thermal conduction in monolayer graphene, *h*-BN, MoS$_2$ and WS$_2$, as well as their three-dimensional (3D) bulk counterparts. Based on full phonon dispersions obtained from *ab initio* simulations, we calculate the in-plane (for monolayer and bulk) and cross-plane (for bulk) ballistic thermal conductance $G_{ball}$ of these materials for the first time. Due to their stronger chemical bonding, graphene (graphite) and *h*-BN generally show higher $G_{ball}$ than MoS$_2$ and WS$_2$ above ~100 K. We then obtain *L*-dependent thermal conductivity of these materials by using a ballistic-diffusive model, which for suspended graphene shows good agreement with recent measurements [32] of *k* vs. *L* up to 9 μm. We also discuss how an estimate of the phonon MFP as a rigorous average of all phonon modes can be condensed to an expression including $G_{ball}$ and the diffusive thermal conductivity $k_{diff}$. We find that kinetic theory can reach the same MFP estimate as long as a correctly averaged phonon group velocity is used. Importantly for anisotropic layered materials, 2D and 1D forms of the kinetic theory should be used for in-plane and out-of-plane transport, respectively. Based on the ballistic-diffusive model, *k* converges to $k_{diff}$ only around $L \sim 100\lambda$, at much larger sizes than commonly assumed. Thus, whether *k* is divergent in low-dimensional materials should be examined beyond ~$100\lambda$ to distinguish from the intrinsic increase of *k* due to the ballistic-to-diffusive transition.

**Phonon Dispersion.** To study the thermal properties of these systems, we begin by calculating the phonon dispersions using a first-principles approach. First, the equilibrium atomic structures are calculated within density functional theory (DFT) using projector-augmented wave (PAW) potentials [33], as implemented in the VASP code [34]. The local density approximation (LDA) is used for the exchange-correlation functional [35]. Then, the interatomic force constants are calculated in a supercell using the frozen-phonon approach [36]. Finally, phonon dispersions in the entire first Brillouin zone (BZ) are calculated by the PHONOPY package [37] interfaced with VASP. Details of these calculations are described in Supplemental Material Section 1.

The atomic lattices of bulk graphite, *h*-BN, and TMDs are shown in Figs. 1(a-c). Graphite tends to crystalize in the AB stacking, while *h*-BN favors the AA' stacking [38,39]. For TMDs, the most common 2H phase [5,40] (similar to the AA' stacking in *h*-BN) is considered here. Figure 2 shows the calculated phonon dispersions along high-symmetry lines in the BZ [see Fig. 1(d)] for bulk graphite, *h*-BN, MoS$_2$ and WS$_2$. The phonon dispersions of their monolayers are shown in Fig. 3. Our calculations are in excellent agreement with available experimental data [41-45] plotted as symbols in Figs. 2 and 3. Among the monolayers shown in Fig. 3, graphene and *h*-BN have six phonon branches due to two atoms in their primitive cells [Figs. 1(a) and 1(b)]; whereas, there are nine phonon branches for monolayer MoS$_2$ and WS$_2$ because of three atoms in their primitive cells [Fig. 1(c)]. For the bulk phonon dispersions shown in Fig. 2, each branch from their monolayer dispersions splits into two branches because two layers form a primitive cell in bulk [Fig. (1)]. The two branches are distinguishable on the ΓMK plane, but they become degenerate on the ALH plane in the BZ (see Supplemental Material Fig. S1). Moreover, the dispersion along the cross-plane direction appears in bulk materials (*e.g.*, Γ-A parts in Fig. 2). More details of the cross-plane dispersion are shown in Supplemental Material Fig. S1.

**Ballistic Thermal Conductance.** From the obtained phonon dispersion, the ballistic thermal conductance ($G_{\text{ball}}$) of a material can be calculated without approximation:

$$G_{\text{ball}} = \frac{1}{2L} \sum_{b,\vec{q}} \hbar \omega_b(\vec{q}) |v_{n,b}(\vec{q})| \frac{\partial f}{\partial T}, \tag{1}$$

where $\hbar$ is the reduced Planck constant; $\vec{q}$ and $\omega_b(\vec{q})$ are phonon wave vector and frequency (*b* denotes different branches), respectively; $L$ and $v_{n,b}(\vec{q}) = \hat{n} \cdot \nabla_{\vec{q}} \omega_b(\vec{q})$ are the material length and phonon group velocity along the heat conducting direction $\hat{n}$ (a normalized vector); $f = 1/[\exp(\hbar\omega_b/k_B T) - 1]$ is the Bose-Einstein distribution; $T$ is temperature and $k_B$ is the Boltzmann constant. The sum in Eq. (1) is over all branches (*b*) and wave vectors ($\vec{q}$) in the entire first BZ. In practical calculations, the sum with respect to $\vec{q}$ is converted to an integral, and the ballistic thermal conductance per unit cross-sectional area (*i.e.*, $G_{\text{ball}}/A$ with $A = WH$, where $W$ and $H$ are the material width and height, respectively) is calculated to give a value independent of size and easy for comparison among different materials. For 2D (monolayer) and 3D (bulk) layered materials, $G_{\text{ball}}/A$ is given, respectively, by

$$\left. \frac{G_{\text{ball}}}{A} \right|_{2D} = \sum_b \frac{1}{8\pi^2 H} \int_{BZ} d\vec{q}^2 \hbar \omega_b(\vec{q}) |v_{n,b}(\vec{q})| \frac{\partial f}{\partial T}, \tag{2a}$$

$$\left. \frac{G_{\text{ball}}}{A} \right|_{3D} = \sum_b \frac{1}{16\pi^3} \int_{BZ} d\vec{q}^3 \hbar \omega_b(\vec{q}) |v_{n,b}(\vec{q})| \frac{\partial f}{\partial T}, \tag{2b}$$

where the integrals are over the entire 2D and 3D BZ, respectively. For monolayers, we take $H = c/2$,

where $c$ is the cross-plane lattice constant of bulk (Fig. 1). For bulk, $v_{n,b}(\vec{q})$ can be $v_{x,b}(\vec{q})$ [or $v_{y,b}(\vec{q})$] and $v_{z,b}(\vec{q})$ to give the in-plane (∥) and cross-plane (⊥) $G_{ball}/A$, respectively (see Supplemental Material Section 3 for the derivation and calculation details).

The calculated ballistic thermal conductance of monolayers, bulk in-plane (∥) and cross-plane (⊥) as a function of temperature are shown in Fig. 4, for graphite, $h$-BN, MoS$_2$ and WS$_2$. For each material, the monolayer $G_{ball}/A$ is higher than that of bulk (∥) at low $T$, but they become almost overlapped at high $T$, because the in-plane phonon dispersions of monolayer and bulk are nearly identical except at low frequencies. For bulk (⊥), its $G_{ball}/A$ is similar to that of bulk (∥) at very low $T$ (< 10 K), but the in-plane (∥) ballistic thermal conductance becomes ~10 times greater at all higher temperatures due to the significant anisotropy of the phonon dispersion.

To put these results in perspective, we also plot the ballistic thermal conductance of silicon, the most widely used semiconductor [14], as a dashed line in each panel of Fig. 4. Above ~100 K, graphene, graphite (∥), $h$-BN monolayer and bulk (∥) show much larger $G_{ball}/A$ than Si [Figs. 4(a) and 4(b)], due to their very strong in-plane covalent bonds which lead to higher phonon group velocity, whereas monolayer and bulk (∥) MoS$_2$ and WS$_2$ show smaller $G_{ball}/A$ than Si at $T$ > 100 K [Figs. 4(c) and 4(d)] due to relatively weak metal-sulfur bonds. All cross-plane (⊥) $G_{ball}/A$ of the four layered materials are lower than that of Si because of their weak van der Waals interactions between layers. The room-temperature $G_{ball}/A$ values and Debye temperatures $\Theta_D$ [45-48] of these materials are listed in Table 1.

In the high $T$ limit, all $G_{ball}/A$ vs. $T$ curves flatten out (saturate) when all phonon modes are fully excited. The temperatures at which $G_{ball}/A$ reach 90% of their maximum value are nearly the same for the monolayer and bulk (∥) of each material. These are ~1150 K, ~1060 K, ~320 K and ~310 K for graphite, $h$-BN, MoS$_2$ and WS$_2$, respectively, which are roughly 60% of their Debye temperature $\Theta_D$ (Table 1). The corresponding temperatures for cross-plane bulk (⊥) are ~780 K, ~1270 K, ~345 K and ~360 K for graphite, $h$-BN, MoS$_2$ and WS$_2$, respectively. The same observation holds approximately for bulk Si as well, where 90% of its ballistic conductance is reached at ~400 K, approximately 60% of its $\Theta_D$ (Table 1).

Interestingly, there is a "bump" in the cross-plane $G_{ball}/A$ of $h$-BN bulk (⊥) before it flattens out, different from other bulk (⊥) curves [Fig. 4(b)]. The physical cause of this "bump" is that the high frequency optical branches (ZO$_1$, ZO$_2$, LO$_1$, LO$_2$) of bulk $h$-BN have noticeably sloped dispersions in the cross-plane direction [see Supplemental Material Fig. S1(f)], leading to *non-zero* phonon velocity $v_z$, unlike the nearly zero $v_z$ in other bulk materials. Thus, as temperature increases and these phonon modes start to contribute to cross-plane conduction, a sharp increase (bump) appears in the $G_{ball}/A$ vs. $T$ curve.

At low temperature, all $G_{ball}/A$ vs. $T$ curves show power law scaling, $\sim T^n$ (Fig. 4). For the monolayers, the power exponent is $n \approx 1.6-1.7$, which is a combined effect of $n = 1.5$ from the quadratic ZA branch and $n = 2$ from the linear TA and LA branches [1,13,15]. The power law only applies to monolayers below ~90 K for graphene and *h*-BN, and below ~40 K for MoS$_2$ and WS$_2$. For bulk (∥), the power exponent is $n \approx 2.9$, which is a combined effect of $n = 2.5$ from the ZA branch (if purely quadratic) [13] and $n = 3$ from linear TA and LA branches (see Supplemental Material Section 8). The reason that the overall $n$ is closer to 3 than 2.5 is that the real ZA branch for bulk has a linear component (not purely quadratic) at very low frequency [49], ultimately leading to $n > 2.5$. The power law for bulk in-plane (∥) is valid only below ~35 K for graphite and *h*-BN, and below ~20 K for MoS$_2$ and WS$_2$. For bulk cross-plane (⊥), the power exponent is $n \approx 2.7-2.9$ with the same physical reason as bulk in-plane (∥), and it applies only below ~10 K for graphite and *h*-BN, and below ~7 K for MoS$_2$ and WS$_2$.

The calculated ballistic thermal conductance is a useful quantity in the investigation of heat conduction limits in nanomaterials and nanoscale devices. It gives the upper limit of heat flow at a given temperature, and also reflects the intrinsic anisotropy of heat conduction in such layered materials. Any measured or calculated thermal *conductivity* ($k$) can be connected to *conductance* ($G$) as $G/A = k/L$, and then compared to $G_{ball}/A$. The quantity $k/L$ must always be less than or equal to $G_{ball}/A$; if this relationship is not satisfied (as for some classical molecular dynamics simulations) the thermal conductivity is unphysical [50]. To put it another way, $k_{ball} = G_{ball}L/A$ represents the *maximum* thermal conductivity a nanoscale device or sample can reach. In addition, the ratio of ($k/L$) to ($G_{ball}/A$) gives the percentage of ballistic conduction that a nanoscale device reaches [15,32]. This approach can also be used to predict thermal conductivity as a function of device length $L$, and to estimate an overall phonon MFP $\lambda$ averaged across all modes.

**Length-Dependent Thermal Conductivity.** We next turn to a discussion of length-dependent thermal conductivity in nanoscale layered materials. In the ballistic regime ($L \ll \lambda$) the thermal conductance approaches a constant, thus the ballistic thermal conductivity $k_{ball} = G_{ball}L/A$ becomes linearly dependent on length $L$. In the diffusive regime ($L \gg \lambda$), the thermal conductivity is typically independent of length $L$ (the case of $k$ divergent with $L$ will be discussed later), becoming a constant $k$

= $k_{\text{diff}}$ (diffusive thermal conductivity). In the diffusive regime the conductance scales inversely with sample length, as a thermal analogy to Ohm's law, $G = kA/L$.

Consequently, in the intermediate ballistic-diffusive (or quasi-ballistic) regime, the effective thermal conductivity should increase with $L$ and gradually converge to $k_{\text{diff}}$; this behavior is similar to the effective mobility dependence on sample length during quasi-ballistic charge transport in short-channel transistors [16,17]. This transition can be captured through a phenomenological ballistic-diffusive (BD) model [15,49]:

$$k(L) \approx \left[ \frac{1}{(G_{\text{ball}}/A)L} + \frac{1}{k_{\text{diff}}} \right]^{-1} \approx \sum_b \left[ \frac{1}{(G_{\text{ball},b}/A)L} + \frac{1}{k_{\text{diff},b}} \right]^{-1}, \quad (3)$$

where the first expression is a "1-color" model, and the second one is a branch-resolved "multi-color" model, taking into account different phonon branch ($b$) contributions. As we will show later, this is essentially a Landauer-like model and it can yield an expression of diffusive thermal conductivity consistent with kinetic theory.

We first apply this model to the thermal conductivity of graphene supported on SiO$_2$. Fig. 5(a) shows the estimated $k$ as a function of $L$ from both experiments [15,51,52] (solid symbols) and Boltzmann transport equation (BTE) calculations [53] (open symbols), at room temperature. By using our calculated $G_{\text{ball}}/A = \Sigma_b G_{\text{ball},b}/A = 4.37$ GWK$^{-1}$m$^{-2}$ for graphene and $k_{\text{diff}} = \Sigma_b k_{\text{diff},b} = 578$ Wm$^{-1}$K$^{-1}$ from BTE simulations, the 1-color model (solid blue line) and 6-color model (solid orange line) as well as its components of each branch (dash-dot lines) are shown in Fig. 5(a). They are in good agreement with experimental data and BTE simulations, indicating that the BD model can provide reasonable $L$-dependent $k$ in this intermediate size regime. In fact, we find that the simple 1-color model is sufficiently good to give almost the same result as the 6-color model, although the latter provides more information about the contribution of each branch. For SiO$_2$-supported graphene, besides the three acoustic branches, the flexural optical (ZO) branch also has a notable contribution, but other optical (TO and LO) branches have negligible contributions (<1%). This occurs because the ZO branch has relatively low phonon frequency [Fig. 3(a)], and hence larger thermal population.

Having shown that the BD model can successfully describe the $k$ change with $L$ in *supported* graphene, next we focus on the $L$-dependent $k$ in *suspended* graphene. Some theoretical studies have predicted that $k$ will diverge with $L$ in isolated low-dimensional (1D and 2D) momentum-conserving systems, *i.e.* $k \sim L^\alpha$ for 1D [20-22] and $k \sim \log(L)$ for 2D like graphene [22-24]. However, these theoretical studies only consider *pure* 1D (2D) systems in which atoms can only move in one (two) dimensions, but atoms in *realistic* 1D (2D) materials like carbon nanotube (graphene) can still move in 3D, leading to properties that pure 1D and 2D systems cannot capture, such as radial breathing mode and flexural mode. It is hence unclear if predicted divergence will happen in realistic materials. In addition, other studies [54-56] have argued that disorder and higher-order three-phonon scattering may eliminate the divergence and, as we will show, no experiments have conclusively observed divergent $k$ yet, in neither 1D nor 2D systems. For supported graphene discussed above, the $\sim\log(L)$ divergence does not appear due to phonon scattering with substrate vibrational modes [57,58], but for suspended graphene it is still an open question. Very recently, Xu *et al*. [32] systematically measured $k$ of suspended graphene as a function of $L$, and their data at $T = 300$ K (green squares) are shown in Fig. 5(b). However, we find that our 1-color BD model [Eq. (3)] can adequately fit these data within the bounds of experimental error, with the fitting parameter $k_{\text{diff}} = 1790$ Wm$^{-1}$K$^{-1}$ [solid line in Fig. 5(b)].

Assuming the divergence of $k \sim \log(L)$ for large $L$, the BD model can be adjusted with a logarithmic term to capture this effect (see Supplemental Material Section 8 for more discussion):

$$k(L) = \left[ \frac{1}{(G_{\text{ball}}/A)L} + \frac{1}{k_{\text{diff}} + k_0 \ln(L/L_0 + 1)} \right]^{-1}, \qquad (4)$$

where $k_0$ and $L_0$ are parameters with units of thermal conductivity and length, respectively. This modified "log" model reproduces the ballistic limit correctly as well, *i.e.* $k \rightarrow (G_{\text{ball}}/A)L$ when $L \rightarrow 0$. Choosing $k_{\text{diff}} = 1590$ Wm$^{-1}$K$^{-1}$, $k_0 = 130$ Wm$^{-1}$K$^{-1}$, and $L_0 = 1.1$ µm, the log model can also fit the data of Xu *et al*. [32] within the bounds of experimental error [dash-dotted line in Fig. 5(b)]. Both models obey the ballistic limit and fit the available data, but they are distinguishable only at $L > 10$ µm, whereas data are available only up to $\sim 9$ µm. In the regime below 10 µm the increase of $k$ with $L$ mainly results from the ballistic-to-diffusive transition. This suggests that the effects of divergent thermal conductivity in suspended graphene will only show up unambiguously in samples longer than $\sim 10$ µm, and that suspended devices and samples shorter than this length can be adequately described with a simple, 1-color BD model.

In Fig. 5(c) we also plot experimental thermal conductivity data [59-62] for suspended single-wall carbon nanotubes (SWCNTs) with similar diameters, as a function of length, at room temperature. There is no systematic measurement of $k$ versus $L$ for SWCNTs, and the few available data do not fall in a single trend, so here we only use the 1-color BD model [Eq. (3)] to fit individual data points (dash-dot lines) and give a "band" (yellow area) to show the $k$ range as $L$ increases [Fig. 5(c)].

As shown by Mingo and Broido [13], SWCNTs have the same ballistic thermal conductance as graphene above ~200 K, i.e. $G_{ball}/A$ = 4.37 GWK$^{-1}$m$^{-2}$ at room temperature. The range of obtained $k_{diff}$ is ~2500−8500 Wm$^{-1}$K$^{-1}$, and thermal conductivity keeps increasing with length up to tens of microns. A more general expression of $k$ including both length and temperature dependence for SWCNTs is provided in our previous study of short channel ($L$ < 100 nm) SWCNT devices [63], where heat conduction is nearly ballistic.

Figure 5(d) summarizes the $L$-dependent thermal conductivities at 300 K for graphite, $h$-BN, MoS$_2$, and Si, including monolayer (1L), bulk in-plane (∥) and bulk cross-plane (⊥) as applicable. These are estimated from the 1-color BD model, given that the measured $k_{diff}$ are known. For suspended graphene, most Raman measurements report $k$ in the range of ~2000−4000 Wm$^{-1}$K$^{-1}$ for $L$ ~ 1−10 μm [1,12], thus we used $k_{diff}$ ≈ 4000 Wm$^{-1}$K$^{-1}$ [64] in the BD model, assuming a convergent $k(L)$. Other $k_{diff}$ used in the BD model are all from experimental measurements: $k_{diff}$ ≈ 2000, 6, 400, 2, 35, 100, 2.5, and 150 Wm$^{-1}$K$^{-1}$ for graphite (∥) [65], graphite (⊥) [65], $h$-BN (∥) [66], $h$-BN (⊥) [67], MoS$_2$ (1L) [68], MoS$_2$ (∥) [69], MoS$_2$ (⊥) [70], and Si [71], respectively (see Table 1). These $k_{diff}$ values are at room temperature, for undoped and isotopically unmodified samples. The layered materials show strong anisotropy in thermal conductivity, and their values span a wide range, more than three orders of magnitude. The estimated $k$ dependence on $L$ given here will be helpful for understanding heat conduction at the length scales where diffusive-ballistic effects take place. We note that convergence to "bulk" thermal conductivity does not occur until 10s of nanometers in the ⊥ direction and 100s of nanometers to microns in the ∥ direction (up to 10 μm in suspended graphene).

**Phonon Mean Free Path and Group Velocity.** Finally, we turn to the estimation of phonon MFP $\lambda$ (averaged across all phonon modes) in terms of the calculated $G_{ball}/A$. We note that the underlying physics of the BD model are rooted in Landauer transport theory, in which the conductance (or conductivity) is given by the ballistic conductance (or conductivity) multiplied by a transmission coefficient [12,14,72]:

$$k(L) \approx \frac{G_{ball}}{A} L \left( \frac{\lambda_{bs}}{L + \lambda_{bs}} \right) = \left[ \frac{1}{(G_{ball}/A)L} + \frac{1}{k_{diff}} \right]^{-1}. \quad (5)$$

The transmission coefficient is $\lambda_{bs}/(L + \lambda_{bs})$, given in terms of the transport length $L$ and an averaged phonon back-scattering MFP, $\lambda_{bs}$. A simple rearrangement of the first expression will reproduce the BD model, meanwhile yielding a relation $\lambda_{bs} = k_{diff}/(G_{ball}/A)$. This relation can be directly derived from the definition of a phonon MFP with all modes weighted properly across the phonon dispersion (see Supplemental Material Section 5). The common phonon MFP $\lambda$ is shorter than the back-scattering MFP $\lambda_{bs}$, and they are related with a factor $\beta_\lambda$ (see Table 2),

$$\lambda = \beta_\lambda \lambda_{bs} = \beta_\lambda \frac{k_{diff}}{G_{ball}/A}. \tag{6}$$

The factor $\beta_\lambda = 1/2$, $2/\pi$, and $3/4$ for 1D, 2D, and 3D materials with *isotropic* thermal properties, respectively [73]. For *anisotropic* layered materials, the in-plane factor is $\beta_\lambda(\parallel) = 2/\pi$, same as isotropic 2D, while the cross-plane factor is $\beta_\lambda(\perp) = 1/2$, same as 1D. The derivation is provided in Supplemental Material Section 5.

As long as reliable $k_{diff}$ from calculations or experiments are available, the overall phonon MFP can be estimated based on Eq. (6), because $G_{ball}/A$ can be calculated without approximations. Taking graphite as an example, using its well-known measured $k_{diff}(T)$ [1,65] and our calculated $G_{ball}(T)/A$, the obtained phonon in-plane and cross-plane MFPs ($\lambda_\parallel$ and $\lambda_\perp$) as a function of temperature are shown as solid lines in Fig. 6(a). They decrease rapidly as $T$ increases and differ by two and one orders of magnitude at low and high $T$, respectively. At room temperature, $\lambda_\parallel \approx 290$ nm and $\lambda_\perp \approx 10$ nm. The latter corresponds to ~30-layers thickness in graphite, which means the cross-plane heat conduction in most few-layer graphene samples [74] is in the quasi-ballistic regime. The advantage of a single MFP averaged over all phonon modes [Eq. (6)] is giving information in a concise way, unlike the use of frequency-dependent MFPs [27-31], and it helps us understand when to make size-effect corrections to thermal conductivity in practical devices.

We also compare our estimates with the classical approach to evaluating the phonon MFP by kinetic theory as $k_{diff} = (1/d)C_v v\lambda$, where $d = 1-3$ is the material dimension, $C_v$ is the heat capacity, and $v$ is the phonon group velocity (typically taken as average sound velocity $v_s$) [75]. The dashed lines in Fig. 6(a) show the $\lambda_\parallel$ and $\lambda_\perp$ of graphite obtained in this way with $v_{s,\parallel} = 15.8$ km/s and $v_{s,\perp} = 2.0$ km/s (see Supplemental Material Section 6). Clearly, the classical $\lambda$ is underestimated compared to our earlier calculation (solid lines) from Eq. (6) because the sound velocity only includes low-frequency phonons and cannot take into account the effect of the quadratic ZA branch, leading to an overestimate of effective phonon group velocity.

To obtain an improved phonon MFP by the classical kinetic theory, the group velocity $v$ should be an average weighted by the heat capacity. By analyzing the integral expressions of $G_{ball}/A$ and $C_v$, we find the averaged group velocity can be simply given by

$$v = \beta_v \frac{G_{\text{ball}}/A}{C_V}, \tag{7}$$

where the velocity factor $\beta_v = 2$, $\pi$, and 4 for isotropic 1D, 2D, and 3D materials. Similar to the MFP factor $\beta_\lambda$, for *anisotropic* layered 3D materials the in-plane $\beta_v = \pi$, same as for isotropic 2D, and the cross-plane $\beta_v = 2$, same as for 1D [see Supplemental Material Section 7 for the derivation of Eq. (7) and $\beta_v$]. Substituting Eq. (7) into the kinetic theory, the phonon MFP is then given by

$$\lambda = \frac{k_{\text{diff}} d}{C_V v} = \frac{d}{\beta_v} \frac{k_{\text{diff}}}{G_{\text{ball}}/A}. \tag{8}$$

We can find $d/\beta_v \equiv \beta_\lambda$ for all cases (see Table 2), thus when the correct $v$ is used, the kinetic theory can give the same estimation of MFP as our model [Eq. (6)]. Figure 6(b) shows the calculated $v$ in terms of Eq. (7) as a function of temperature for graphene, graphite ($\parallel$), and graphite ($\perp$). In the whole temperature range, the maximum in-plane averaged phonon velocity for graphite and graphene is $v_\parallel \sim 8$ km/s, about half of the averaged sound velocity (15.8 km/s). For graphite cross-plane, the averaged $v_\perp \approx 345$ m/s at room temperature, almost six times smaller than 2 km/s from the sound velocity average, and it reaches 2 km/s only below $\sim$30 K, *i.e.*, the $T$ range where the sound velocity is valid. The calculated average phonon velocities based on Eq. (7) for $h$-BN, MoS$_2$, and WS$_2$, as a function of temperature, are shown in Supplemental Material Fig. S4.

After demonstrating the BD model and kinetic theory can be reconciled in the MFP estimation, we apply Eq. (6) to other interesting materials whose measured $k_{\text{diff}}$ are available. Figure 6(c) shows the estimated $\lambda$ at 300 K are $\sim$3, 9, 10, 25, 70, 76, 90, 100, 290, 580 nm for $h$-BN ($\perp$), MoS$_2$ ($\perp$), graphite ($\perp$), MoS$_2$ (1L), $h$-BN ($\parallel$), MoS$_2$ ($\parallel$), graphene (on SiO$_2$), Si, graphite ($\parallel$), and graphene (suspended), respectively. The $k_{\text{diff}}$ used are summarized in Table 1.

By comparing the obtained $\lambda$ with $k(L)$ in Fig. 5(d), we can find that $k$ reaches $\sim$85% of $k_{\text{diff}}$ at $L \approx 10\lambda$ and $\sim$90% at 16$\lambda$. At larger device dimensions $k$ increases slowly and reaches $\sim$98% of $k_{\text{diff}}$ (essentially fully diffusive) at $L = 100\lambda$, which is much longer than the commonly assumed sample sizes. The entire ballistic-to-diffusive transition takes more than two orders of magnitude in sample size (length), from $L \approx \lambda$ to 100$\lambda$ to complete. This also indicates that any obtained $k$ increase in this size regime is likely to arise from the ballistic-to-diffusive transition, and whether $k$ is divergent in low-dimensional materials should be examined in samples with $L > 100\lambda$.

We note that Eq. (6) gives an MFP including contributions from all phonon modes, but some modes (*e.g.*, optical) have smaller contributions to thermal conductivity and short MFPs (<10 nm for in-plane and <1 nm for cross-plane). This implies that the modes contributing significantly to heat conduction should have MFPs longer than the value given by Eq. (6). Indeed, by analyzing the cumulative thermal conductivity as a function of MFP, it was found that phonons with MFPs longer than 1 μm and 100 nm contribute ~40% to the total thermal conductivity for Si [27-31] and cross-plane graphite [76] at 300 K, respectively. These length scales correspond to a "median" MFP of ~10$\lambda$ based on our definition from Eq. (6). Our approach provides a single, rigorous MFP with all phonon modes weighted properly (see Supplemental Material Section 5), which should be used in the kinetic theory and to understand quasi-ballistic thermal transport. From Fig. 5(d), when $L < \lambda$, heat conduction enters the ballistic regime ($k$ linear with $L$) as expected. When $L$ is shorter than the "median" MFP suggested above (~10$\lambda$), $k$ is still a strong function of $L$, but is not in the ballistic regime, implying that this is a characteristic length scale below which the size effects (not just ballistic transport) take place.

In conclusion, we calculated the ballistic thermal conductance $G_{ball}$ of graphene and graphite, $h$-BN, MoS$_2$, and WS$_2$ based on full phonon dispersions obtained from *ab initio* simulations. We obtained the size-dependent thermal conductivity $k$ with a simple ballistic-diffusive model verified against Boltzmann transport calculations. We showed that a single phonon MFP $\lambda$ can be simply determined by $G_{ball}$ and $k_{diff}$, and that kinetic theory can reach the same result if a proper phonon group velocity is used. The MFP $\lambda$ provides nanoscale heat flow information in a concise way, which is of great utility for simple and rapid estimates in practical nanoscale devices. Based on the calculated $\lambda$ and our approach, we find that $k$ converges to 90% $k_{diff}$ at $L \sim 16\lambda$, and full convergence is only achieved for sample dimensions $L > 100\lambda$, much longer than commonly assumed. This also implies that to verify divergent $k$ in low-dimensional systems, simulations and measurements should be performed for $L > 100\lambda$.

**Acknowledgement**

We acknowledge valuable discussions with Chen Si for the first-principle calculations of phonon dispersions. This work was in part supported by the Presidential Early Career (PECASE) award W911NF-13-1-0471 from the Army Research Office, and National Science Foundation (NSF) awards DMREF 1534279 and EFRI 1542883 (ZL and EP). YL and WD acknowledge the support of the Ministry of Science and Technology of China (grant 2016YFA0301001), and the National

Natural Science Foundation of China (grant 11674188 and 11334006). LL acknowledges support from the Department of Energy, Basic Energy Sciences, Materials Sciences and Engineering Division for work done at Oak Ridge National Laboratory.

**Supplemental Material**

Details of first-principles calculations, additional plots of phonon dispersions, derivations of heat capacity, ballistic thermal conductance, diffusive thermal conductivity, averaged phonon mean free path and group velocity for different cases, as well as complimentary discussion are provided. This material is available free of charge via the Internet at [URL to be inserted by Publisher].

**REFERENCES**

[1] E. Pop, V. Varshney, and A. K. Roy, *Thermal properties of graphene: Fundamentals and applications*, MRS Bull. **37**, 1273 (2012).

[2] A. A. Balandin, *Thermal properties of graphene and nanostructured carbon materials*, Nat. Mater. **10**, 569 (2011).

[3] B. Huang, Q. M. Yan, Z. Y. Li, and W. H. Duan, *Towards graphene nanoribbon-based electronics*, Front. Phys. China **4**, 269 (2009).

[4] G. Fiori, F. Bonaccorso, G. Iannaccone, T. Palacios, D. Neumaier, A. Seabaugh, S. K. Banerjee, and L. Colombo, *Electronics based on two-dimensional materials*, Nat. Nanotechnol. **9**, 768 (2014).

[5] Q. H. Wang, K. Kalantar-Zadeh, A. Kis, J. N. Coleman, and M. S. Strano, *Electronics and optoelectronics of two-dimensional transition metal dichalcogenides*, Nat. Nanotechnol. **7**, 699 (2012).

[6] A. K. Geim and I. V. Grigorieva, *Van der Waals heterostructures*, Nature **499**, 419 (2013).

[7] S. J. Choi, B. K. Kim, T. H. Lee, Y. H. Kim, Z. Li, E. Pop, J. J. Kim, J. H. Song, and M. H. Bae, *Electrical and Thermoelectric Transport by Variable Range Hopping in Thin Black Phosphorus Devices*, Nano Lett. **16**, 3969 (2016).

[8] D. G. Cahill *et al.*, *Nanoscale thermal transport. II. 2003–2012*, Appl. Phys. Rev. **1**, 011305 (2014).

[9] E. Pop, *Energy Dissipation and Transport in Nanoscale Devices*, Nano Res. **3**, 147 (2010).

[10] S. Islam, Z. Li, V. E. Dorgan, M.-H. Bae, and E. Pop, *Role of Joule Heating on Current Saturation and Transient Behavior of Graphene Transistors*, IEEE Electron Device Lett. **34**, 166 (2013).

[11] F. Lian, J. P. Llinas, Z. Li, D. Estrada, and E. Pop, *Thermal conductivity of chirality-sorted carbon nanotube networks*, Appl. Phys. Lett. **108**, 103101 (2016).

[12]    Y. Xu, Z. Li, and W. Duan, *Thermal and Thermoelectric Properties of Graphene*, Small **10**, 2182 (2014).

[13]    N. Mingo and D. A. Broido, *Carbon Nanotube Ballistic Thermal Conductance and Its Limits*, Phys. Rev. Lett. **95**, 096105 (2005).

[14]    C. Jeong, S. Datta, and M. Lundstrom, *Full dispersion versus Debye model evaluation of lattice thermal conductivity with a Landauer approach*, J. Appl. Phys. **109**, 073718 (2011).

[15]    M.-H. Bae, Z. Li, Z. Aksamija, P. N. Martin, F. Xiong, Z.-Y. Ong, I. Knezevic, and E. Pop, *Ballistic to diffusive crossover of heat flow in graphene ribbons*, Nat. Commun. **4**, 1734 (2013).

[16]    M. S. Shur, *Low ballistic mobility in submicron HEMTs*, IEEE Electron Device Lett. **23**, 511 (2002).

[17]    J. Wang and M. Lundstrom, *Ballistic transport in high electron mobility transistors*, IEEE Trans. Electron Devices **50**, 1604 (2003).

[18]    X. Du, I. Skachko, A. Barker, and E. Y. Andrei, *Approaching ballistic transport in suspended graphene*, Nat. Nanotechnol. **3**, 491 (2008).

[19]    K. Vonklitzing, G. Dorda, and M. Pepper, *New Method for High-Accuracy Determination of the Fine-Structure Constant Based on Quantized Hall Resistance*, Phys. Rev. Lett. **45**, 494 (1980).

[20]    B. W. Li and J. Wang, *Anomalous heat conduction and anomalous diffusion in one-dimensional systems*, Phys. Rev. Lett. **91**, 044301 (2003).

[21]    T. Prosen and D. K. Campbell, *Momentum conservation implies anomalous energy transport in 1D classical lattices*, Phys. Rev. Lett. **84**, 2857 (2000).

[22]    G. Basile, C. Bernardin, and S. Olla, *Momentum conserving model with anomalous thermal conductivity in low dimensional systems*, Phys. Rev. Lett. **96**, 204303 (2006).

[23]    A. Lippi and R. Livi, *Heat conduction in two-dimensional nonlinear lattices*, J. Stat. Phys. **100**, 1147 (2000).

[24]    L. Wang, B. B. Hu, and B. W. Li, *Logarithmic divergent thermal conductivity in two-dimensional nonlinear lattices*, Phys. Rev. E **86**, 040101(R) (2012).

[25]    J. A. Johnson, A. A. Maznev, J. Cuffe, J. K. Eliason, A. J. Minnich, T. Kehoe, C. M. S. Torres, G. Chen, and K. A. Nelson, *Direct Measurement of Room-Temperature Nondiffusive Thermal Transport Over Micron Distances in a Silicon Membrane*, Phys. Rev. Lett. **110**, 025901 (2013).

[26]    Y. S. Ju and K. E. Goodson, *Phonon scattering in silicon films with thickness of order 100 nm*, Appl. Phys. Lett. **74**, 3005 (1999).

[27]    K. Esfarjani, G. Chen, and H. T. Stokes, *Heat transport in silicon from first-principles calculations*, Phys. Rev. B **84**, 085204 (2011).

[28]    W. Li, N. Mingo, L. Lindsay, D. A. Broido, D. A. Stewart, and N. A. Katcho, *Thermal conductivity of diamond nanowires from first principles*, Phys. Rev. B **85**, 195436 (2012).

[29]    F. Yang and C. Dames, *Mean free path spectra as a tool to understand thermal*

*conductivity in bulk and nanostructures*, Phys. Rev. B **87**, 035437 (2013).

[30] K. T. Regner, D. P. Sellan, Z. H. Su, C. H. Amon, A. J. H. McGaughey, and J. A. Malen, *Broadband phonon mean free path contributions to thermal conductivity measured using frequency domain thermoreflectance*, Nat. Commun. **4**, 1640 (2013).

[31] C. Jeong, S. Datta, and M. Lundstrom, *Thermal conductivity of bulk and thin-film silicon: A Landauer approach*, J. Appl. Phys. **111**, 093708 (2012).

[32] X. F. Xu *et al.*, *Length-dependent thermal conductivity in suspended single-layer graphene*, Nat. Commun. **5**, 3689 (2014).

[33] P. E. Blöchl, *Projector Augmented-Wave Method*, Phys. Rev. B **50**, 17953 (1994).

[34] G. Kresse and J. Furthmüller, *Efficient iterative schemes for ab initio total-energy calculations using a plane-wave basis set*, Phys. Rev. B **54**, 11169 (1996).

[35] J. P. Perdew and Y. Wang, *Accurate and Simple Analytic Representation of the Electron-Gas Correlation-Energy*, Phys. Rev. B **45**, 13244 (1992).

[36] M. T. Yin and M. L. Cohen, *Theory of Lattice-Dynamical Properties of Solids - Application to Si and Ge*, Phys. Rev. B **26**, 3259 (1982).

[37] A. Togo, F. Oba, and I. Tanaka, *First-principles calculations of the ferroelastic transition between rutile-type and CaCl(2)-type SiO(2) at high pressures*, Phys. Rev. B **78**, 134106 (2008).

[38] R. S. Pease, *Crystal Structure of Boron Nitride*, Nature **165**, 722 (1950).

[39] G. Constantinescu, A. Kuc, and T. Heine, *Stacking in Bulk and Bilayer Hexagonal Boron Nitride*, Phys. Rev. Lett. **111**, 036104 (2013).

[40] M. Chhowalla, H. S. Shin, G. Eda, L. J. Li, K. P. Loh, and H. Zhang, *The chemistry of two-dimensional layered transition metal dichalcogenide nanosheets*, Nat. Chem. **5**, 263 (2013).

[41] R. Nicklow, H. G. Smith, and Wakabaya.N, *Lattice-Dynamics of Pyrolytic-Graphite*, Phys. Rev. B **5**, 4951 (1972).

[42] H. Yanagisawa, T. Tanaka, Y. Ishida, M. Matsue, E. Rokuta, S. Otani, and C. Oshima, *Analysis of phonons in graphene sheets by means of HREELS measurement and ab initio calculation*, Surf. Interface Anal. **37**, 133 (2005).

[43] M. Mohr, J. Maultzsch, E. Dobardzic, S. Reich, I. Milosevic, M. Damnjanovic, A. Bosak, M. Krisch, and C. Thomsen, *Phonon dispersion of graphite by inelastic x-ray scattering*, Phys. Rev. B **76**, 035439 (2007).

[44] J. Serrano, A. Bosak, R. Arenal, M. Krisch, K. Watanabe, T. Taniguchi, H. Kanda, A. Rubio, and L. Wirtz, *Vibrational properties of hexagonal boron nitride: Inelastic X-ray scattering and ab initio calculations*, Phys. Rev. Lett. **98**, 095503 (2007).

[45] N. Wakabayashi, H. G. Smith, and R. M. Nicklow, *Lattice-Dynamics of Hexagonal Mos2 Studied by Neutron-Scattering*, Phys. Rev. B **12**, 659 (1975).

[46] T. Tohei, A. Kuwabara, F. Oba, and I. Tanaka, *Debye temperature and stiffness of carbon and boron nitride polymorphs from first principles calculations*, Phys. Rev. B **73**, 064304 (2006).

[47] C. H. Ho, C. S. Wu, Y. S. Huang, P. C. Liao, and K. K. Tiong, *Temperature dependence*

*of energies and broadening parameters of the band-edge excitons of Mo1-xWxS2 single crystals*, J. Phys. Condens. Mat. **10**, 9317 (1998).

[48]    C. Kittel, *Introduction to Solid State Physics* (John Wiley & Sons, Inc, 2005), Eighth edn.

[49]    R. Prasher, *Thermal boundary resistance and thermal conductivity of multiwalled carbon nanotubes*, Phys. Rev. B **77**, 075424 (2008).

[50]    M. M. Sadeghi, M. T. Pettes, and L. Shi, *Thermal transport in graphene*, Solid State Commun. **152**, 1321 (2012).

[51]    J. H. Seol *et al.*, *Two-Dimensional Phonon Transport in Supported Graphene*, Science **328**, 213 (2010).

[52]    Z. Li, M.-H. Bae, and E. Pop, *Substrate-supported thermometry platform for nanomaterials like graphene, nanotubes, and nanowires*, Appl. Phys. Lett. **105**, 023107 (2014).

[53]    Calculations follow the method described in Ref. 51, including the use of an empirical potential to describe the atomic interactions.

[54]    P. L. Garrido, P. I. Hurtado, and B. Nadrowski, *Simple one-dimensional model of heat conduction which obeys Fourier's law*, Phys. Rev. Lett. **86**, 5486 (2001).

[55]    N. Mingo and D. A. Broido, *Length dependence of carbon nanotube thermal conductivity and the "problem of long waves"*, Nano Lett. **5**, 1221 (2005).

[56]    Z. L. Wang, D. W. Tang, X. H. Zheng, W. G. Zhang, and Y. T. Zhu, *Length-dependent thermal conductivity of single-wall carbon nanotubes: prediction and measurements*, Nanotechnology **18**, 475714 (2007).

[57]    J. Chen, G. Zhang, and B. W. Li, *Substrate coupling suppresses size dependence of thermal conductivity in supported graphene*, Nanoscale **5**, 532 (2013).

[58]    Z.-Y. Ong and E. Pop, *Effect of substrate modes on thermal transport in supported graphene*, Phys. Rev. B **84**, 075471 (2011).

[59]    C. H. Yu, L. Shi, Z. Yao, D. Y. Li, and A. Majumdar, *Thermal conductance and thermopower of an individual single-wall carbon nanotube*, Nano Lett. **5**, 1842 (2005).

[60]    E. Pop, D. Mann, Q. Wang, K. E. Goodson, and H. J. Dai, *Thermal conductance of an individual single-wall carbon nanotube above room temperature*, Nano Lett. **6**, 96 (2006).

[61]    Z. L. Wang *et al.*, *Length-dependent thermal conductivity of an individual single-wall carbon nanotube*, Appl. Phys. Lett. **91**, 123119 (2007).

[62]    Q. W. Li, C. H. Liu, X. S. Wang, and S. S. Fan, *Measuring the thermal conductivity of individual carbon nanotubes by the Raman shift method*, Nanotechnology **20**, 145702 (2009).

[63]    M. M. Shulaker *et al.*, *Carbon Nanotube Circuit Integration up to Sub-20 nm Channel Lengths*, Acs Nano **8**, 3434 (2014).

[64]    Here the used value $k_{diff}$ = 4000 W/m/K and the fitted value $k_{diff}$ = 1790 W/m/K in Fig. 5(b) are not contradictory, in fact they represent the most likely range of suspended graphene $k_{diff}$ reported in measurements.

[65]    C. Y. Ho, R. W. Powell, and P. E. Liley, *Thermal Conductivity of the Elements*, J. Phys. Chem. Ref. Data **1**, 279 (1972).

[66] E. K. Sichel, R. E. Miller, M. S. Abrahams, and C. J. Buiocchi, *Heat-Capacity and Thermal-Conductivity of Hexagonal Pyrolytic Boron-Nitride*, Phys. Rev. B **13**, 4607 (1976).

[67] A. Simpson and A. D. Stuckes, *The Thermal Conductivity of Highly Orientated Pyrolytic Boron Nitride*, J Phys. C: Solid St. Phys. **4**, 1710 (1971).

[68] R. S. Yan *et al.*, *Thermal Conductivity of Monolayer Molybdenum Disulfide Obtained from Temperature-Dependent Raman Spectroscopy*, ACS Nano **8**, 986 (2014).

[69] J. Liu, G. M. Choi, and D. G. Cahill, *Measurement of the anisotropic thermal conductivity of molybdenum disulfide by the time-resolved magneto-optic Kerr effect*, J. Appl. Phys. **116**, 233107 (2014).

[70] C. Muratore *et al.*, *Cross-plane thermal properties of transition metal dichalcogenides*, Appl. Phys. Lett. **102**, 081604 (2013).

[71] A. D. McConnell and K. E. Goodson, *Thermal conduction in silicon micro- and nanostructures*, Annual Reviews of Heat Transfer **14**, 129 (2005).

[72] Y. Xu, J. S. Wang, W. H. Duan, B. L. Gu, and B. W. Li, *Nonequilibrium Green's function method for phonon-phonon interactions and ballistic-diffusive thermal transport*, Phys. Rev. B **78**, 224303 (2008).

[73] C. Jeong, R. Kim, M. Luisier, S. Datta, and M. Lundstrom, *On Landauer versus Boltzmann and full band versus effective mass evaluation of thermoelectric transport coefficients*, J. Appl. Phys. **107**, 023707 (2010).

[74] Y. K. Koh, M.-H. Bae, D. G. Cahill, and E. Pop, *Heat Conduction across Monolayer and Few-Layer Graphenes*, Nano Lett. **10**, 4363 (2010).

[75] J. M. Ziman, *Electrons and Phonons* (Oxford University Press, 1960).

[76] Z. Y. Wei, J. K. Yang, W. Y. Chen, K. D. Bi, D. Y. Li, and Y. F. Chen, *Phonon mean free path of graphite along the c-axis*, Appl. Phys. Lett. **104**, 081903 (2014).

**FIGURES**

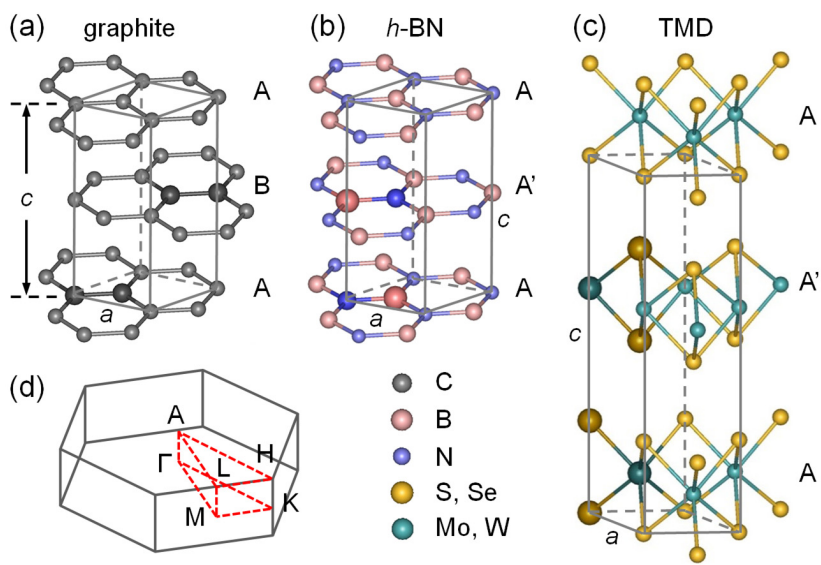

**FIG. 1.** Atomic structures for graphite with AB stacking (a), *h*-BN with AA' stacking (b), and TMD with 2H-phase (c), where three atomic layers are plotted for each of them. Gray lines indicate primitive cells for their bulk, and lattice constants *a* and *c* are labeled. Unrepeatable atoms within the primitive cells are highlighted by bigger and darker colored balls: 4 atoms for graphite and *h*-BN, 6 atoms for TMDs. (d) Schematic of their Brillouin zone with high-symmetry lines highlighted. For their 2D monolayers, the material "thickness" is generally defined as *c*/2, and the 2D BZ is simply a regular hexagon [see inset of Fig. 3(a)].

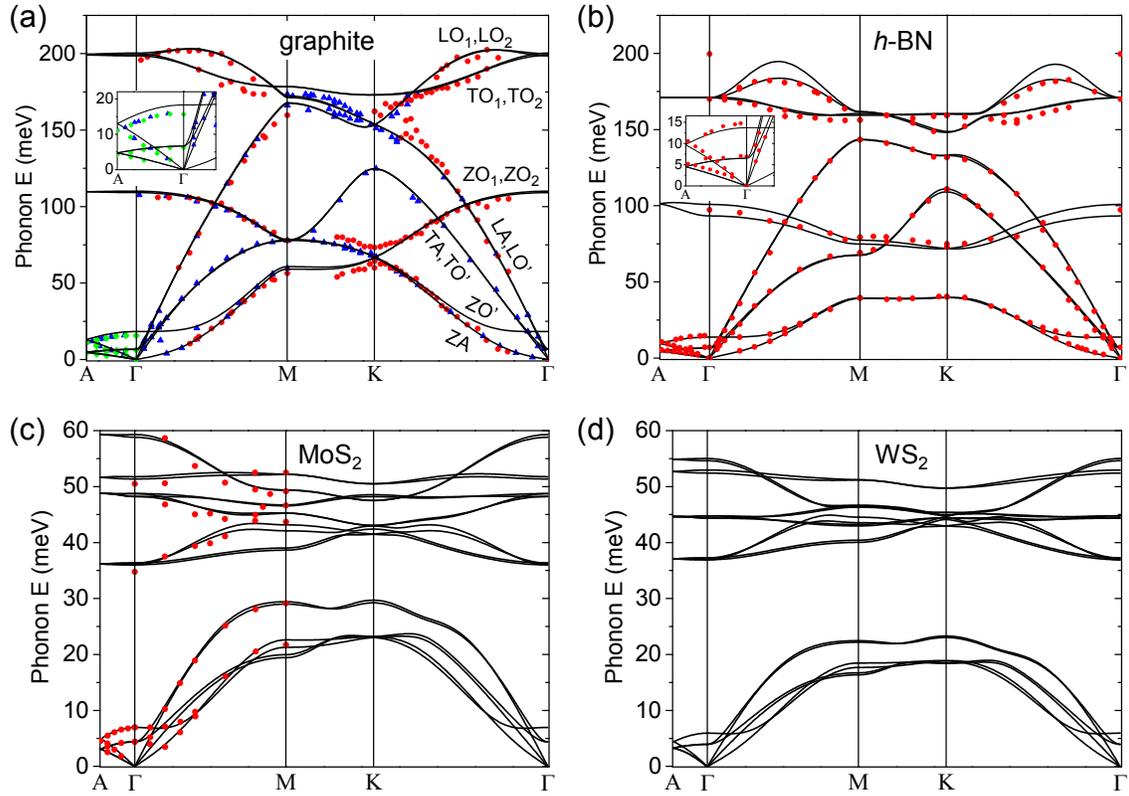

**FIG. 2.** Phonon dispersions in *bulk* layered materials: (a) graphite, (b) *h*-BN, (c) MoS$_2$ and (d) WS$_2$. Insets in (a) and (b) are zoom-in dispersions along the cross-plane direction [shown in greater detail in Supplemental Material Fig. S1(b-c)]. Lines are the calculated phonon dispersions and symbols are available experimental data. (Green diamonds [41], red circles [42], and blue triangles [43] for graphite, and other experimental data shown for *h*-BN [44] and MoS$_2$ [45].)

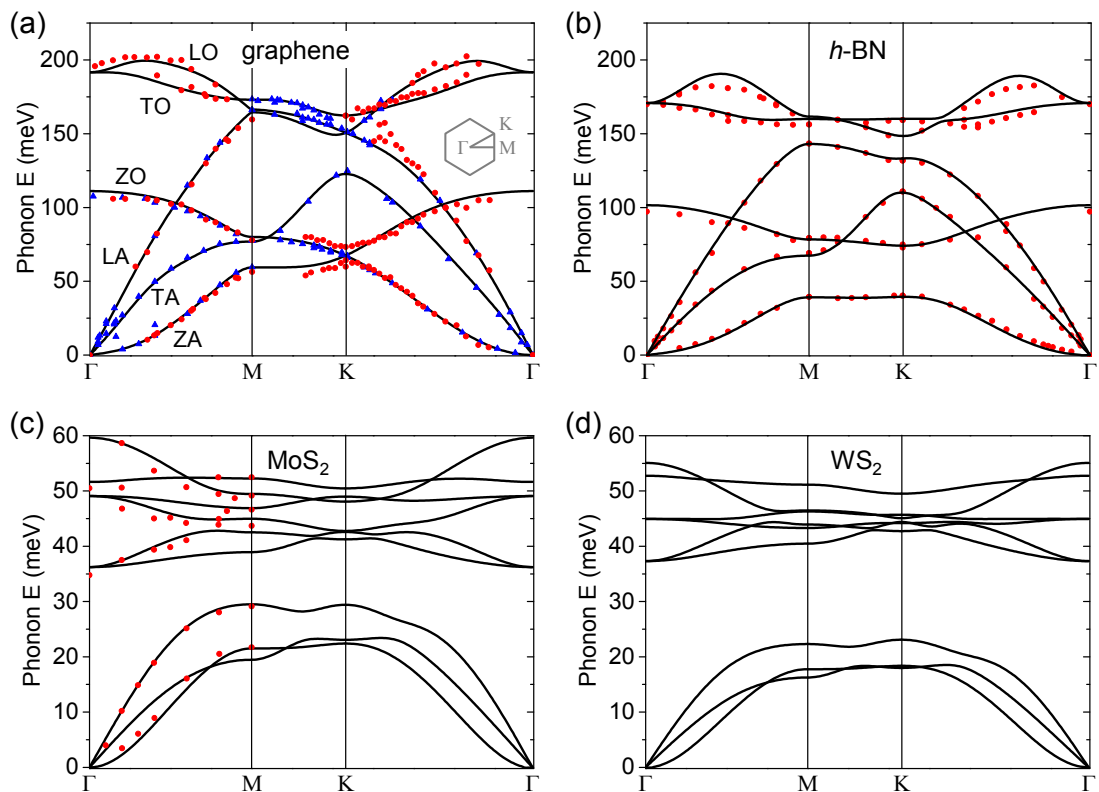

**FIG. 3.** Phonon dispersions in *monolayer* materials: (a) graphene, (b) *h*-BN, (c) MoS$_2$ and (d) WS$_2$. Lines are the calculated phonon dispersions and symbols are available experimental data for bulk in-plane modes. (Data for graphene are represented as red circles [42] and blue triangles [43]; experimental data are also shown for *h*-BN[44] and MoS$_2$ [45].) The 2D Brillouin zone for monolayers is displayed as the inset in (a).

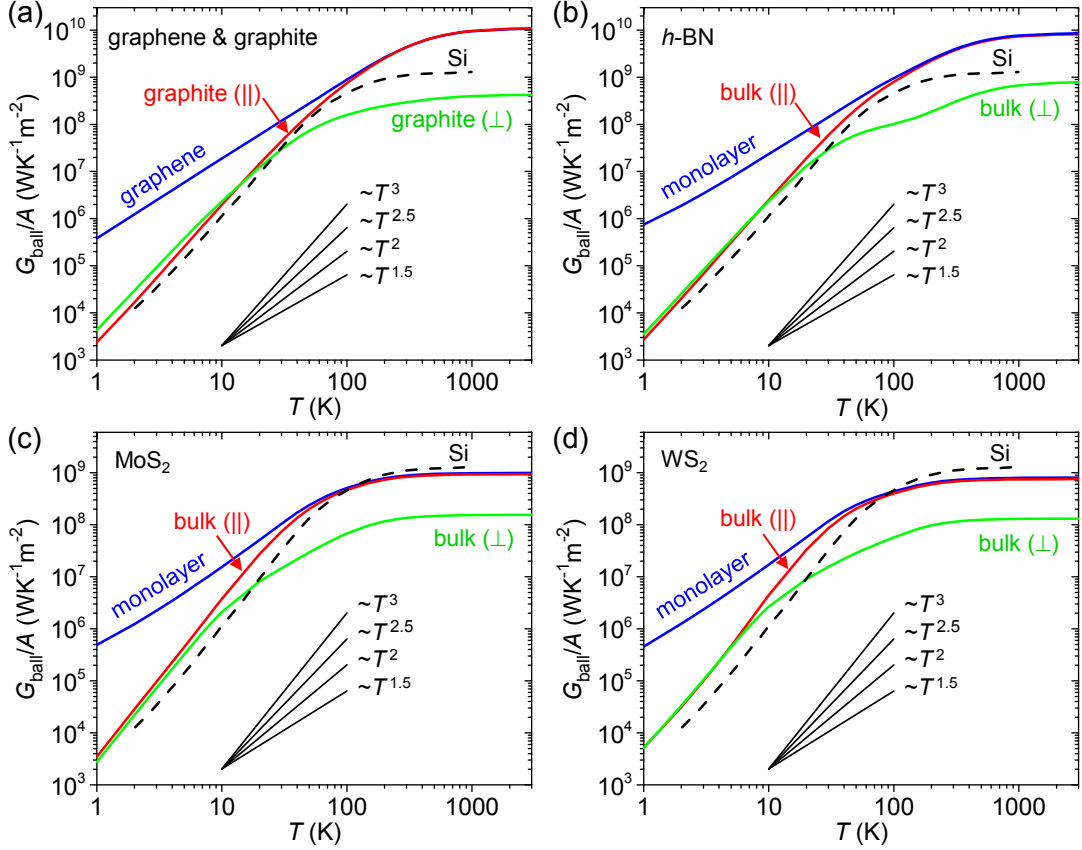

**FIG. 4.** Calculated ballistic thermal conductance per cross-sectional area ($G_{ball}/A$) as a function of temperature for: (a) graphene and graphite, (b) $h$-BN, (c) $MoS_2$ and (d) $WS_2$. Each panel includes the ballistic conductance limit of the respective monolayer, bulk in-plane ($\parallel$), and bulk cross-plane ($\perp$), as well as that of silicon [14] (dashed lines) for comparison.

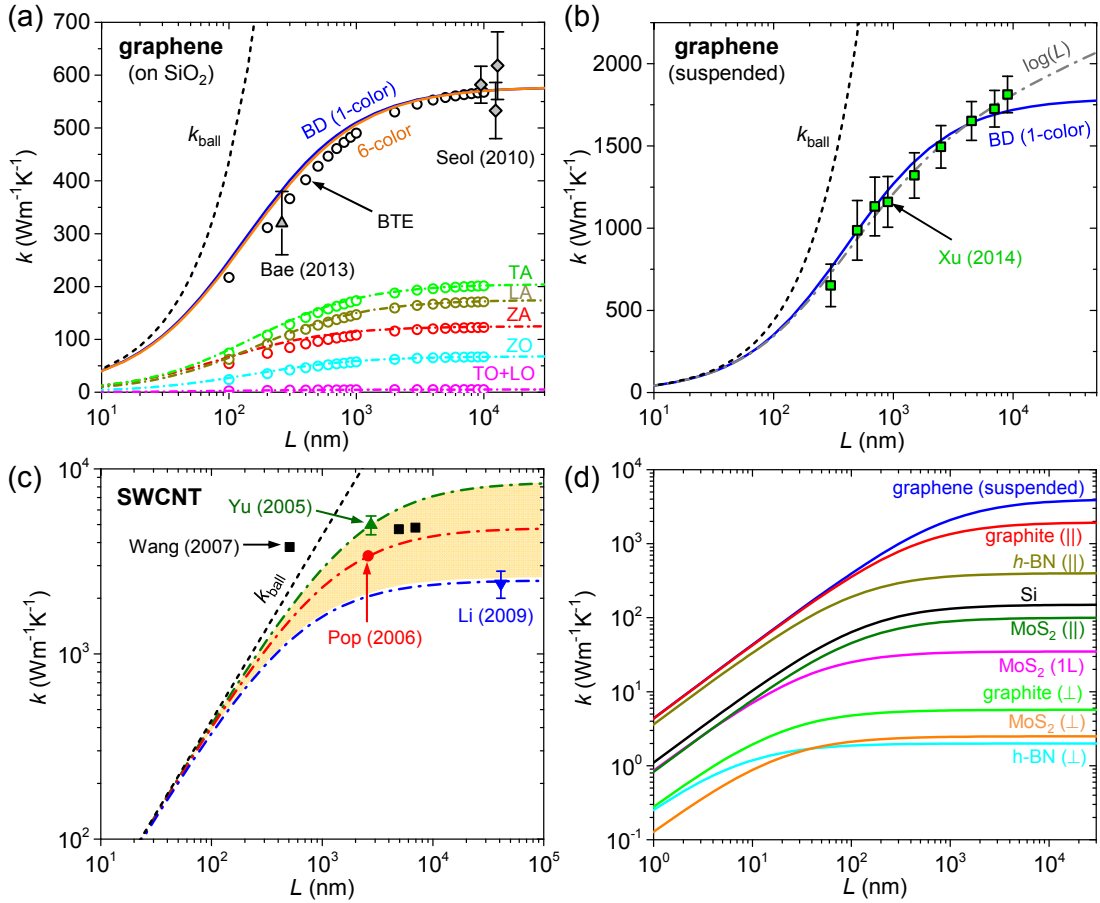

**FIG. 5.** Size effects in the dependence of thermal conductivity $k$ on sample size $L$ at room temperature: (a) Graphene supported on SiO$_2$. The 1-color (solid blue line) and 6-color (solid orange line) BD models from Eq. (3) show good agreement with BTE calculations (open circles) and experimental data (triangle [15] and diamonds [51]). The contributions of individual phonon branches from the 6-color model and BTE are also shown. (b) Suspended graphene. Both the 1-color BD model (solid blue line) from Eq. (3) and the modified "log" model (dash-dot line) from Eq. (4) can fit the available experimental data [32]. (c) Suspended single-wall carbon nanotubes (SWCNTs). Experimental data of Yu *et al*. [59], Pop *et al*. [60], Wang *et al*. [61], and Li *et al*. [62] are plotted. One of the black squares shows a data point that is most likely unphysical, as it lies above the ballistic limit $k_{\text{ball}}$. Dash-dotted lines are the 1-color BD model fitted to individual data points, and the yellow band shows the overall range of confidence based on the available data. (d) Expected size dependence of thermal conductivity for various layered materials and their monolayers, compared with Si.

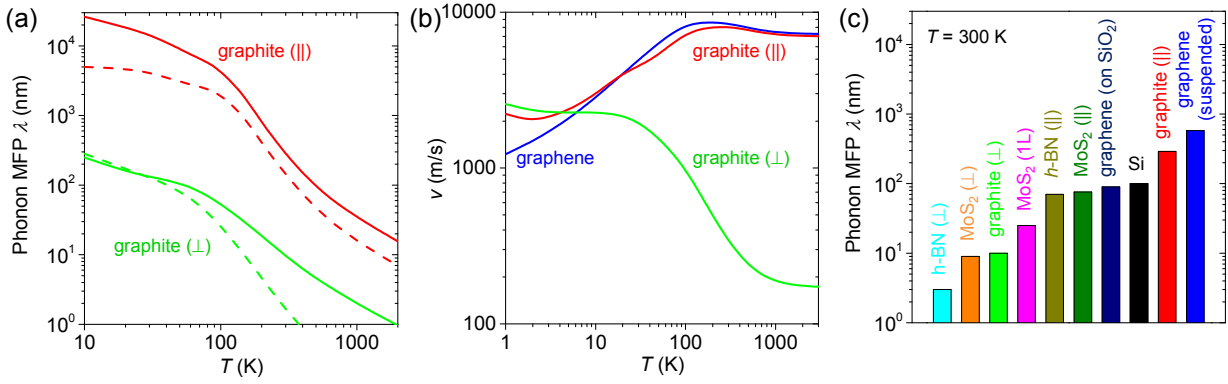

**FIG. 6.** (a) Phonon MFP $\lambda$ as a function of temperature for graphite ($\|$) and ($\perp$). The kinetic theory using the classical sound velocity underestimates $\lambda$ (dash lines) compared to the results from Eq. (6) (solid lines). (b) Averaged phonon group velocity as a function of temperature obtained from Eq. (7) for graphene, graphite ($\|$), and graphite ($\perp$). Supplemental Material Fig. S4 shows similar plots for $h$-BN, MoS$_2$, and WS$_2$. (c) Room temperature phonon MFP $\lambda$ for different materials obtained from Eq. (6) using the calculated $G_{\text{ball}}$ and experimental $k_{\text{diff}}$ from the literature.

**TABLE 1.** Calculated room-temperature $G_{ball}/A$ for layered materials of interest, available Debye temperatures $\Theta_D$ and diffusive thermal conductivities from the literature. The values listed in parenthesis correspond to $\Theta_D$ in the low-temperature limit for MoS$_2$ and WS$_2$. Superscripts correspond to the respective reference sources, not to exponents.

| | Debye temperature $\Theta_D$ (K) | $G_{ball}/A$ (GWK$^{-1}$m$^{-2}$) | | | $k_{diff}$ (Wm$^{-1}$K$^{-1}$) | | |
|---|---|---|---|---|---|---|---|
| | | monolayer | bulk (∥) | bulk (⊥) | monolayer (suspended) | bulk (∥) | bulk (⊥) |
| graphite | 1930 [46] | 4.37 | 4.34 | 0.294 | 2000−4000 [1] | 2000 [65] | 6 [65] |
| h-BN | 1740 [46] | 3.85 | 3.65 | 0.291 | -- | 400 [66] | 2 [67] |
| MoS$_2$ | 570 (253) [45,47] | 0.88 | 0.83 | 0.135 | 35 [68] | 85−110 [69] | 2 [69], 2.5 [70] |
| WS$_2$ | (210) [47] | 0.72 | 0.67 | 0.114 | -- | -- | -- |
| Si | 645 [48] | | 1.11 [14] | | | 150 [71] | |

**TABLE 2.** Values of $\beta_\lambda$ in Eq. (6), $\beta_v$ in Eq. (7), and $d$ in the kinetic theory $k_{diff} = (1/d)C_v v \lambda$ for different cases, where $d = \beta_\lambda \beta_v$. We note that the listed values for layered 3D are *not* applicable to materials whose *in-plane* thermal properties are *anisotropic*, such as phosphorus.

| | 1D | 2D isotropic | 3D isotropic | layered 3D | |
|---|---|---|---|---|---|
| | | | | in-plane | cross-plane |
| $\beta_\lambda$ | 1/2 | 2/$\pi$ | 3/4 | 2/$\pi$ | 1/2 |
| $\beta_v$ | 2 | $\pi$ | 4 | $\pi$ | 2 |
| $d$ | 1 | 2 | 3 | 2 | 1 |